\documentclass{eas}
\usepackage{graphicx}
\newcommand{\arcsec}{\mbox{$^{\prime\prime}$}}

\newcommand{\Msun}{\mbox{\,$M_\odot$}}

\newcommand{\AU}{\mbox{\footnotesize AU}}

\TitreGlobal{The Physics of Evolved Stars}
\begin{document}

\title{Panchromatic Imaging and Spectroscopic Observations of the Mass Ejections of RY Scuti} 
\author{Robert D. Gehrz}\address{Minnesota Institute for Astrophysics, 
School of Physics and Astronomy, 116 Church Street, S.E., University of Minnesota, Minneapolis, Minnesota 55455, USA}
\author{Nathan Smith}\address{Steward Observatory, University of Arizona, 933 N. Cherry Ave, Tucson, AZ 85721, USA}
\author{Dinesh Shenoy}\sameaddress{1}
\vspace{-4mm}
\begin{abstract}
We describe recent panchromatic imaging and spectroscopic studies of the supergiant, mass-transferring, binary star RY Scuti, which is in a brief transitional phase to become a Wolf-Rayet star and a stripped-envelope supernova.
\end{abstract}
\maketitle
\vspace{-4mm}
\section{Introduction}
Massive stellar systems like RY Scuti, 1.8 $\pm$ 0.2 kpc distant (Smith et al. \cite{sm02}), contribute metals and heavy elements to the Inter-Stellar Medium (ISM) during their Post Main Sequence (PMS) evolution and supernova (SN) explosions. Fossil structures that develop during their PMS mass-loss phase can profoundly affect the geometry of the SN remnant.  Moreover, RY Scuti is a special system because it is a massive eclipsing binary caught in the brief phase of Roche lobe overflow, where one of the massive stars is being stripped of its H envelope to yield a Wolf-Rayet star and eventually a Type Ibc SN. RY Scuti consists of an 8\Msun\ O9.7Ibpe primary and a 30\Msun\ B0.5I secondary (Grundstrom et al. \cite{gr01}, Smith et al. \cite{sm11}). The companion star is surrounded by an accretion disk.  Infrared (IR), optical, and radio images show that material with excess angular momentum has expanded in the equatorial plane to yield a toroidal nebula, viewed nearly edge-on, that has been spatially resolved into two components (see Figures 1 and 2). The observations presented here delineate the current distribution of circumstellar matter produced by multiple ejections during the last few hundred years, and therefore offer important constraints on the fundamental parameters of binary mass transfer in SN progenitors. An outer dust torus, with a dust mass of $\simeq$\ 10$^{-6}$\Msun\ and a diameter of 3600 \AU\ (2\arcsec), is expanding at a rate consistent with an ejection 250 years ago (Smith et al. \cite{sm11}). An inner torus seen in H$\alpha$ emission (Figure 2) absorbs the star's ionizing radiation and shields the outer torus, allowing the dust to survive.  The inner torus has a gas mass of 0.003 \Msun\ and a diameter of 2300 \AU\ (1.3\arcsec).  It was ejected 125 years ago (Smith et al. \cite{sm11}).

\begin{figure}[htb!]
\includegraphics[width=6cm]{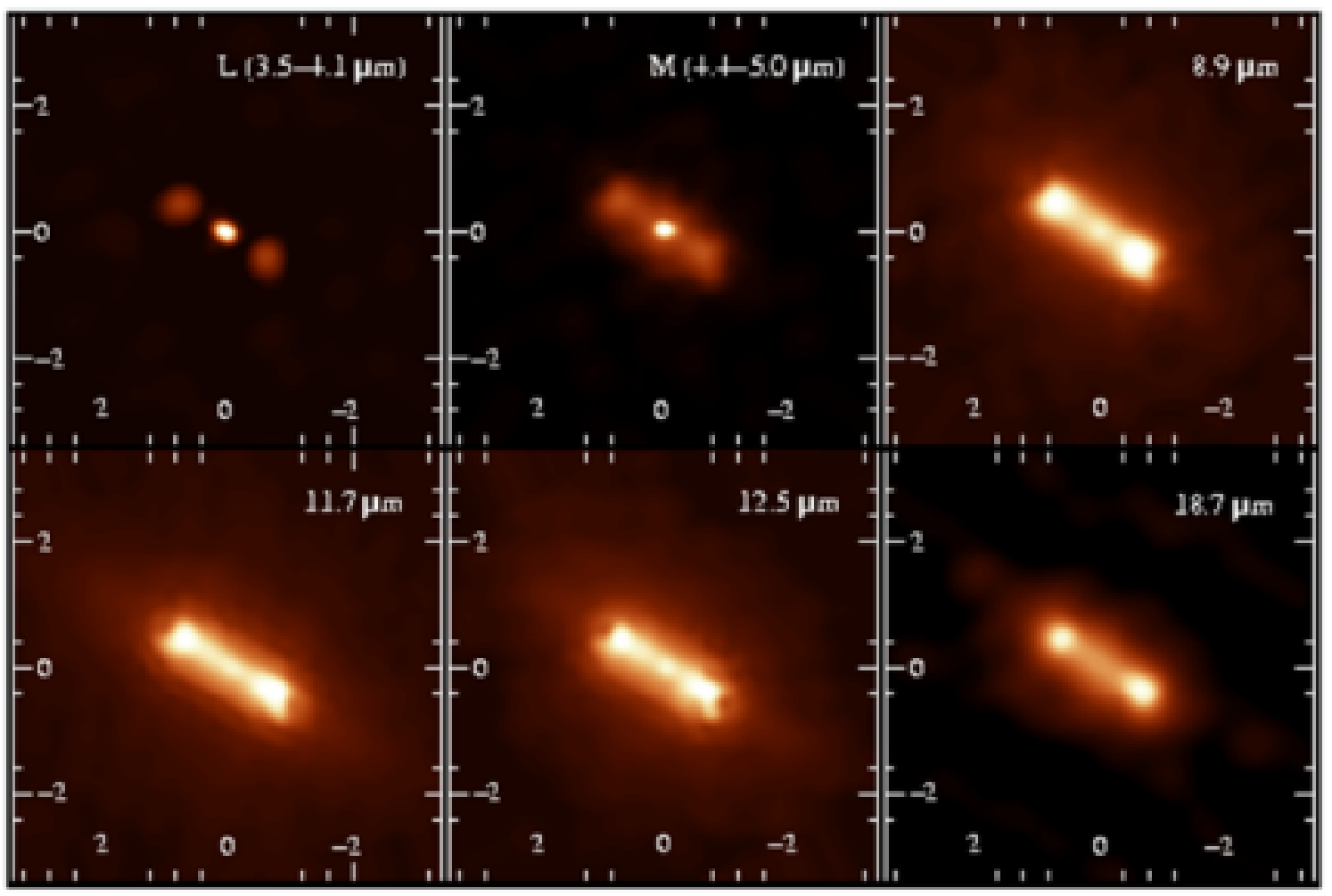}
\qquad
\includegraphics[width=6cm]{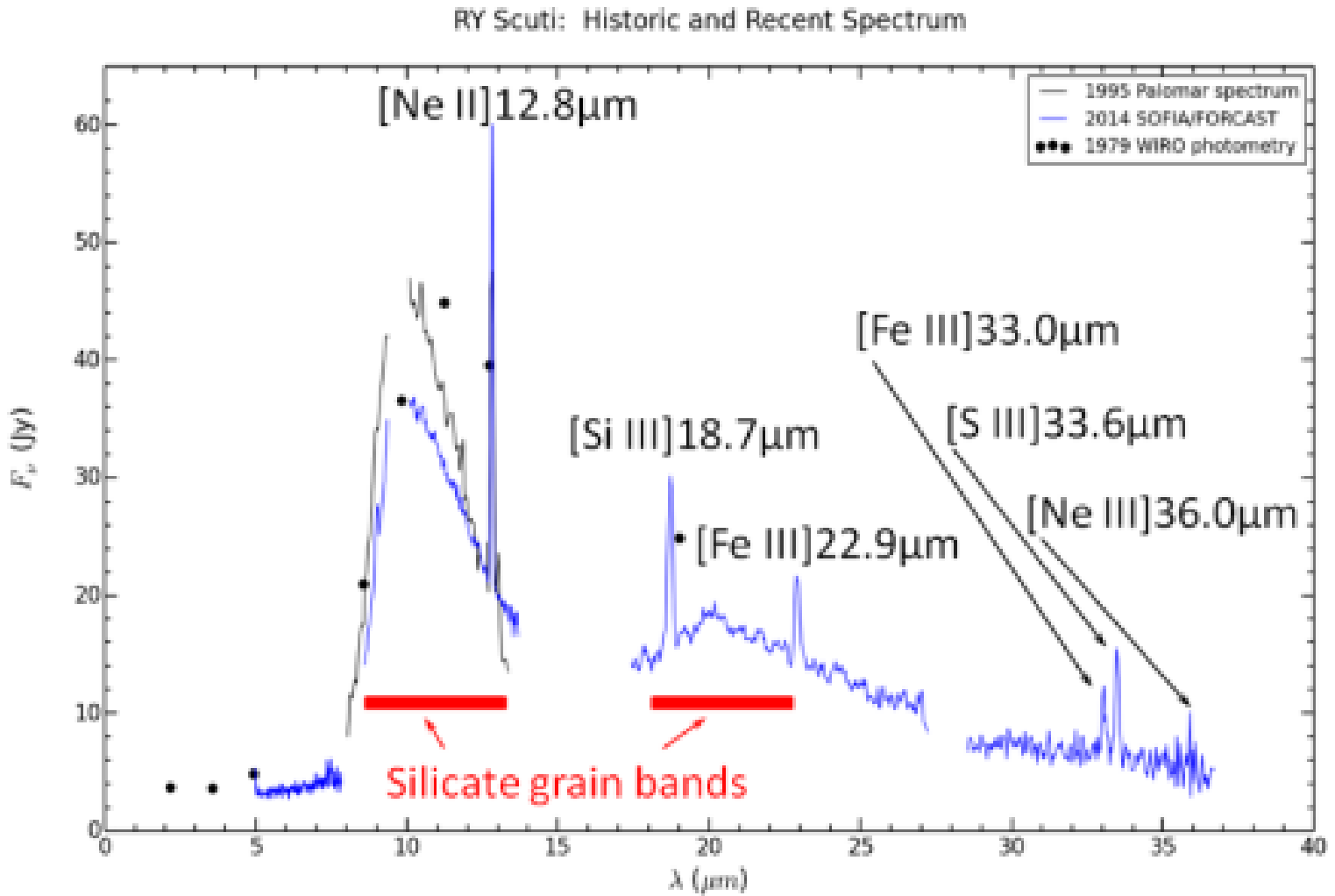}
\caption{{\bf Left:} IR 3.5-18.7 $\mu$m Keck images (Gehrz et al. \cite{ge01}) show the dust torus.  The Rayleigh-Jeans tail of the central hot stellar component dies out at long wavelengths. {\bf Right:} SOFIA FORCAST (Young et al. \cite{yo12}) spectrum is dominated by silicate emission the dust and forbidden line emission from the gas (Smith et al. \cite{sm15}).}
\end{figure}
\begin{figure}[htb!]
\includegraphics[width=6cm]{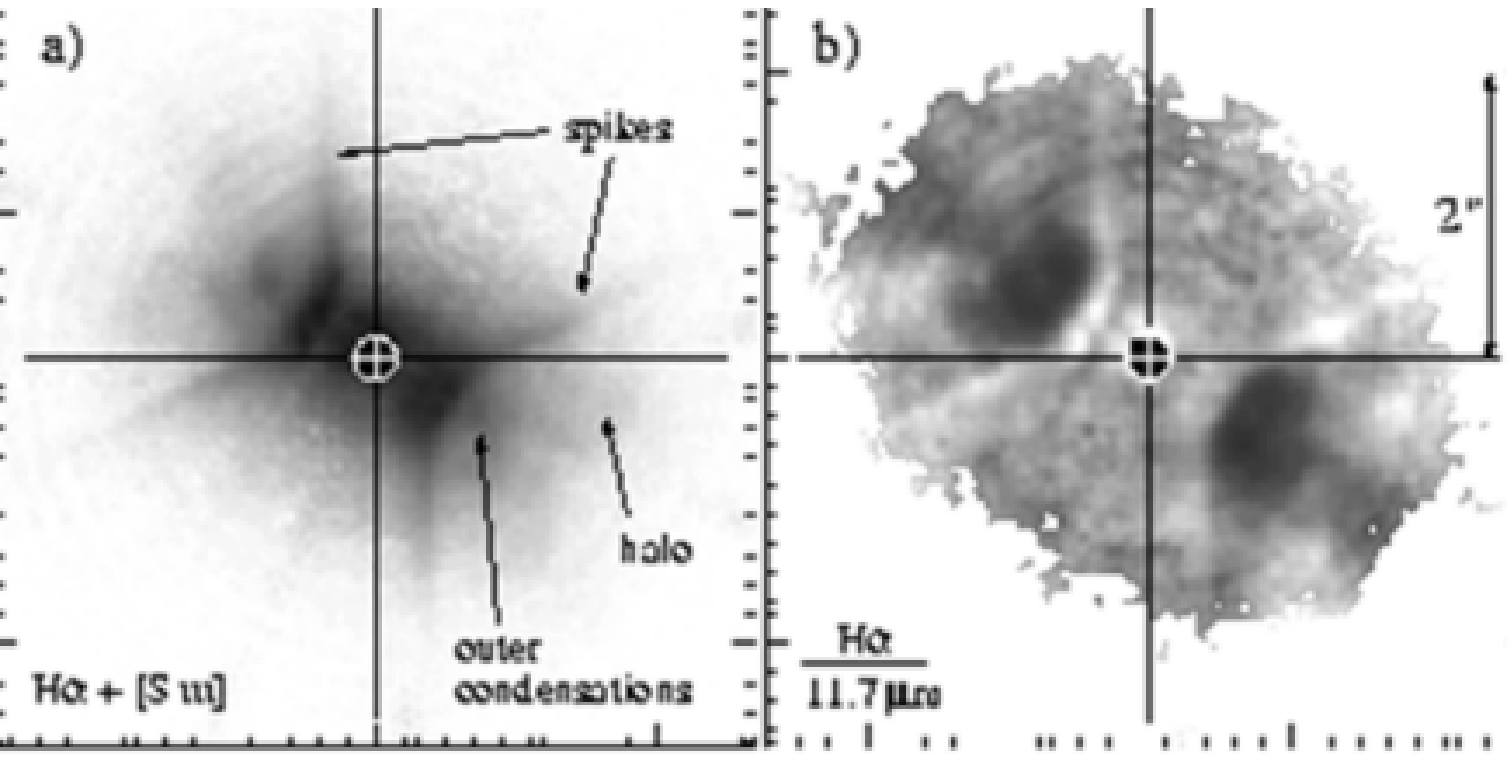}
\qquad
\includegraphics[width=6cm]{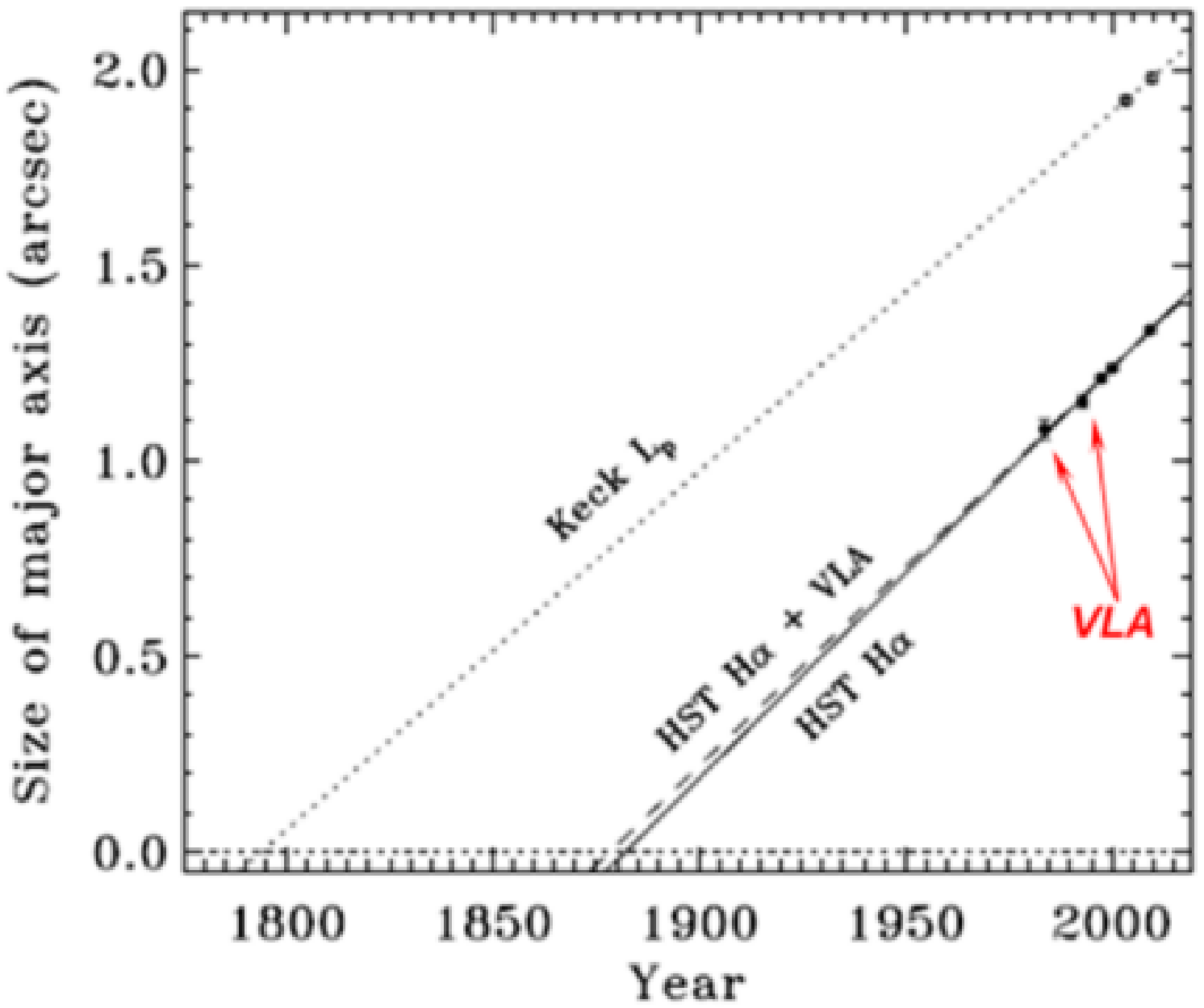}
\caption{{\bf Left:} Superposition of the Keck IR, HST H$\alpha$, and VLA images shows the spatial separation of the gas and dust tori (Smith et al. \cite{sm01}), and {\bf Right:} reveals their separate ejection dates (Smith et al. \cite{sm11}).}
\end{figure}

\vspace{-4mm}

\end{document}